\documentclass[onecolumn,doublespacing]{mn}
\usepackage{graphics}

\newcommand{\be}{\begin{equation}}
\newcommand{\ee}{\end{equation}}
\newcommand{\bea}{\begin{eqnarray}}
\newcommand{\eea}{\end{eqnarray}}
\newcommand{\bc}{\begin{center}}

\newcommand{\ec}{\end{center}}
\def\spose#1{\hbox to 0pt{#1\hss}}
\newcommand{\lta}{\mathrel{\spose{\lower 3pt\hbox{$\mathchar"218$}}
     \raise 2.0pt\hbox{$\mathchar"13C$}}}
\newcommand{\gta}{\mathrel{\spose{\lower 3pt\hbox{$\mathchar"218$}}
     \raise 2.0pt\hbox{$\mathchar"13E$}}}

\def\H0{$H_0= 100~h~$km\,s$^{-1}$\,Mpc$^{-1}$}

\newif\ifAMStwofonts

\ifoldfss
  \ifCUPmtlplainloaded \else
    \NewTextAlphabet{textbfit} {cmbxti10} {}
    \NewTextAlphabet{textbfss} {cmssbx10} {}
    \NewMathAlphabet{mathbfit} {cmbxti10} {} 
    \NewMathAlphabet{mathbfss} {cmssbx10} {} 
  \fi
  \ifAMStwofonts
    \ifCUPmtlplainloaded \else
      \NewSymbolFont{upmath} {eurm10}
      \NewSymbolFont{AMSa} {msam10}
      \NewMathSymbol{\upi}     {0}{upmath}{19}
      \NewMathSymbol{\umu}     {0}{upmath}{16}
      \NewMathSymbol{\upartial}{0}{upmath}{40}
      \NewMathSymbol{\leqslant}{3}{AMSa}{36}
      \NewMathSymbol{\geqslant}{3}{AMSa}{3E}

      \let\leq=\leqslant 
      \let\geq=\geqslant \let\ge=\geqslant
    \fi
  \fi
\fi 

\ifnfssone
  \newmathalphabet{\mathit}
  \addtoversion{normal}{\mathit}{cmr}{m}{it}
  \addtoversion{bold}{\mathit}{cmr}{bx}{it}
  \newmathalphabet{\mathbfit} 
  \addtoversion{normal}{\mathbfit}{cmr}{bx}{it}
  \addtoversion{bold}{\mathbfit}{cmr}{bx}{it}
  \newmathalphabet{\mathbfss} 
  \addtoversion{normal}{\mathbfss}{cmss}{bx}{n}
  \addtoversion{bold}{\mathbfss}{cmss}{bx}{n}
  \ifAMStwofonts
    \ifCUPmtlplainloaded \else
and
your
      \UseAMStwoboldmath
      \makeatletter
      \new@mathgroup\upmath@group
      \define@mathgroup\mv@normal\upmath@group{eur}{m}{n}
      \define@mathgroup\mv@bold\upmath@group{eur}{b}{n}
      \edef\UPM{\hexnumber\upmath@group}
      \new@mathgroup\amsa@group
      \define@mathgroup\mv@normal\amsa@group{msa}{m}{n}
      \define@mathgroup\mv@bold\amsa@group{msa}{m}{n}
      \edef\AMSa{\hexnumber\amsa@group}
      \makeatother
      \mathchardef\upi="0\UPM19
      \mathchardef\umu="0\UPM16
      \mathchardef\upartial="0\UPM40
      \mathchardef\leqslant="3\AMSa36
      \mathchardef\geqslant="3\AMSa3E

      \let\leq=\leqslant 
      \let\geq=\geqslant \let\ge=\geqslant
    \fi
  \fi
\fi 

\ifnfsstwo
  \DeclareMathAlphabet{\mathbfit}{OT1}{cmr}{bx}{it}
  \SetMathAlphabet\mathbfit{bold}{OT1}{cmr}{bx}{it}
  \DeclareMathAlphabet{\mathbfss}{OT1}{cmss}{bx}{n}
  \SetMathAlphabet\mathbfss{bold}{OT1}{cmss}{bx}{n}

  \ifAMStwofonts
    \ifCUPmtlplainloaded \else
      \DeclareSymbolFont{UPM}{U}{eur}{m}{n}
      \SetSymbolFont{UPM}{bold}{U}{eur}{b}{n}
      \DeclareSymbolFont{AMSa}{U}{msa}{m}{n}
      \DeclareMathSymbol{\upi}{0}{UPM}{"19}
      \DeclareMathSymbol{\umu}{0}{UPM}{"16}
      \DeclareMathSymbol{\upartial}{0}{UPM}{"40}
      \DeclareMathSymbol{\leqslant}{3}{AMSa}{"36}
      \DeclareMathSymbol{\geqslant}{3}{AMSa}{"3E}

      \let\leq=\leqslant 
      \let\geq=\geqslant \let\ge=\geqslant
    \fi
  \fi
\fi 

\ifCUPmtlplainloaded \else
  \ifAMStwofonts \else 
    \def\upi{\pi}
    \def\umu{\mu}
    \def\upartial{\partial}
  \fi
\fi

\title{The mass and temperature functions in a moving barrier model.}
\author[A.Del Popolo]
  {A. Del Popolo,$^1$$^,$$^2$$^,$$^3$\\
  $^1$ Dipartimento di Matematica, Universit\`{a} Statale di Bergamo,
  via dei Caniana, 2,  24127, Bergamo, ITALY \\
     $^2$ Feza G\"ursey Institute, P.O. Box 6 \c Cengelk\"oy, Istanbul,
     Turkey\\
     $3$  Bo$\breve{g}azi$\c{c}i University, Physics Department,
     80815 Bebek, Istanbul, Turkey
}
\date{Accepted ???
      Received 2000 July 24;
      in original form ???}

\pagerange{\pageref{firstpage}--\pageref{lastpage}} \pubyear{2000}

\begin{document}

\maketitle

\label{firstpage}

\begin{abstract}

In this paper, I use the extension of the excursion set model of Sheth \& Tormen (2002) and 
the barrier shape obtained in Del Popolo \& Gambera (1998) to calculate the unconditional 
halo mass function, and the conditional mass function in several cosmological models. I show that the 
barrier obtained in Del Popolo \& Gambera (1998), which takes account of 
tidal interaction
between proto-haloes, is a better description of the mass functions than the spherical collapse and 
is in good agreement with numerical simulations (Tozzi \& Governato 1998, and Governato et al. 1999). 
The results are also in good agreement with those obtained 
by Sheth \& Tormen (2002), only slight differences are observed expecially at the low mass end. 
I moreover calculate, and compare with simulations, 
the temperature function obtained by means of the mass functions previously calculated and also 
using an improved version of the M-T relation,  
which accounts for the fact that massive clusters accrete matter quasi-continuously, 
and finally taking account of the tidal interaction with neighboring clusters.
Even in this case the discrepancy between the Press-Schecter predictions and 
simulations is considerably reduced. 
\end{abstract}



\section{Introduction}

In the most promising cosmological scenarios,
structure formation is traced back to the hierarchical growth of primordial
Gaussian density fluctuations originated from quantum fluctuations
(Guth \& Pi 1982; Hawking 1982; Starobinsky 1982; Bardeen et al. 1986 -  hereafter BBKS). 
Starting from these fluctuations, collapsed, virialized dark matter haloes condensed out. Within these haloes
gas cools and stars form (White \& Reese 1977; White \& Frenk 1991; Kauffmann et al. 1999). 
So, the structure of dark matter haloes is of fundamental importance in the
study of the formation and evolution of galaxies and
clusters of galaxies. From the theoretical point of view, the structure of
dark matter haloes can be studied both analytically and numerically.
An analytical model that has achieved a wide popularity is the Press-Schecter (1974) (hereafter PS) formula, which 
allows one to compute good approximations to the mass function. PS and Bond et al. (1991) gave a detailed description of the PS statistics together with an understanding of its dynamical basis. In the PS and Bond et al (1991) 
papers, the authors described how the statistical properties of the initial density field, assumed to be Gaussian, 
together with the spherical collapse model, could be used to derive an estimate of the number density of 
collapsed dark matter haloes at later times, the so called universal ``unconditional" mass function. 
Lacey and Cole (1993), showed how 
the model could be extended to estimate the merging rate of small objects to form larger ones, thus leading 
to the possibility of estimating the ``conditional" mass function of sub-haloes within parent haloes. Mo \& White (1996) 
applied the model to compute an approximation to the spatial clustering of dark haloes. 

Although the analytical framework of the PS model has been greatly refined and extended (as testified by the previous 
cited papers), it is well known that the PS mass function, while qualitatively correct, disagrees with the results of 
N-body simulations. In particular, the PS formula overestimates the abundance of haloes near the characteristic mass 
$M_{\ast}$ and underestimates the abundance in the high mass tail (Efstathiou et al. 1988; White, Efstathiou \& Frenk 1993; Lacey \& Cole 1994; Tozzi \& Governato 1998; Gross et al. 1998; Governato et al. 1999). The quoted discrepancy 
is not surprising since the PS model, as any other analytical model, should make several assumptions to get simple 
analytical predictions. As previously reported, the main assumptions that the PS model combines are the simple physics 
of the spherical collapse model with the assumption that the initial fluctuations were Gaussian and small. 
On average, initially denser regions collapse before less dense ones, which means that, at any given epoch, 
there is a critical density, $\delta_c(z)$, which must be exceeded if collapse is to occur. 
In the spherical collapse model, this critical density does not depend on the mass of the collapsed object. Taking account of the effects of asphericity and 
tidal interaction with neighbors, Del Popolo \& Gambera (1998) and Sheth, Mo \& Tormen (2001) (hereafter SMT),
using a parametrization of the ellipsoidal collapse, showed that the threshold is mass dependent, and in particular that of the set of 
objects that collapse at the same time, the less massive ones must initially have been denser than the more massive, since the less massive ones would have had to hold themselves together against stronger tidal forces. 
%
%
In the second hand, the Gaussian nature of the fluctuation field means that a good approximation to the number density 
of bound objects that have mass $m$ at time $z$ is given by considering the barrier crossing statistics of many independent and uncorrelated random walks, where the barrier shape $B(m,z)$, is connected to the collapse threshold. 
Simply changing the barrier shape, SMT showed that it is possible to incorporate the ``quoted effects" \footnote{Namely that in the case of objects collapsing at the same time, the less massive regions must initially have been denser than the more massive ones.}
%
%
in the excursion set approach. 
Moreover, using the shape of the modified barrier in the excursion set approach, it is possible to obtain a good fit to the universal halo mass function. \footnote{Note that at present there is no good numerical test of analytic predictions for the low mass tail of the mass function.} 
As previously reported, the excursion set approach allows one to calculate good approximations to several 
important quantities, such as the ``unconditional" and ``conditional" mass functions. 
Sheth \& Tormen (2002) (hereafter ST) provided formulas to calculate these last quantities starting from the shape of the 
barrier. They also showed that while the 
``unconditional" and ``conditional" mass function is in good agreement with results from numerical simulations, neither the 
constant nor the moving barrier models (barrier obtained from non-spherical collapse) were able to describe the simulations results at small lookback times, in the case of the rescaled (in terms of $\nu$) ``conditional" mass function. The reason for this discrepancy is probably due to the excursion set approach's neglect of correlations between scales (Peacock \& Heavens 1990; Bond et al. 1991; ST) or to the too simple parametrization of the ellipsoidal collapse outlined in SMT. 
 
In the present paper, I'll use the barrier shape obtained in Del Popolo \& Gambera (1998), obtained from the parametrization of the nonlinear collapse discussed in that paper, together with the results of ST in order to study the ``unconditional" and ``conditional" mass function. 
Finally, I'll calculate the temperature function for a CDM model by means of the mass functions previously obtained and using an improved version of the M-T relation obtained by Voit (2000), which accounts for the fact that massive clusters accrete matter quasi-continuously, and consequently that the M-T relation evolves, with time, more modestly than what expected in previous models (top-hat model) and taking account of the tidal interaction with neighboring clusters.

The reasons that motivates this study are several:\\
a) to study the effects of a barrier different from that used by ST on both ``unconditional" and ``conditional" mass function, as proposed even by ST
; \\
%
%
b) to study how well ST formulas really do work for several barrier shapes;\\
c) to test if the discrepancies between the temperature function and simulations, observed in several papers (e.g. Governato et al. 1999) are reduced using the mass function and the M-T relation obtained.

The paper is organized as follows: in Sect. ~2, I calculate the ``unconditional" and ``conditional" mass functions.
In Sect. ~3, I introduce a model for the mass-temperature (M-T) relation and calculate the temperature function.
Sect. ~4 and 5 are devoted to results and to conclusions, respectively.

\section{The barrier model and the ``unconditional" and ``conditional" mass functions}

Following Sheth \& Tormen (1999) notation, if $f(m,\delta) dm$ denotes the fraction of mass that is contained in collapsed haloes that have mass in the range $m$-$m+dm$, at redshift $z$, and $\delta(z)$ the redshift dependent overdensity, 
the associated ``unconditional" mass function is:
\begin{equation}
n(m,\delta)dm=\frac{\overline{\rho}}{m} f(m,\delta) dm
\end{equation}
where $\overline{\rho}$ is the background density.
If $f(m,\delta_1|M,\delta_0)$ denotes the fraction of the mass of a halo $M$ at $z_0$ that was in subhaloes of mass $m$ at $z_1$, ($z_1>z_0$), the ``conditional" mass function is:
\begin{equation}
N(m,\delta _{1}|M,\delta _{o})dm\equiv \frac{M}{m}f(m,\delta _{1}|M,\delta _{o})dm
\end{equation}

In the excursion set approach, the average comoving number density of haloes of mass $m$ 
the universal or ``unconditional" mass function, $n(m,z)$, is given by:
\begin{equation}
n(m,z)=\frac{\overline{\rho}}{m^{2}}\frac{d\log{\nu }}{d\log m}\nu f(\nu )
\label{eq:universal}
\end{equation}
(Bond et al. 1991), where $\overline{\rho}$ is the background density, $\nu=\left(\frac{\delta_{\rm c}(z)}{\sigma(m)}\right)^2$ is 
the ratio between the critical overdensity required for collapse in the spherical model, $\delta_{\rm c}(z)$, to the r.m.s. density fluctuation $\sigma(m)$, on the scale $r$ of the initial size of the object $m$. The function $\nu f(\nu)$ is obtained by computing the distribution of first crossings, $f(\nu) d \nu$, of a barrier $B(\nu)$, by independent, uncorrelated Brownian motion random walks. The mass function can be thus calculated once a shape for the barrier is given and the power spectrum is known. In the case of spherical collapse, characterized by a constant barrier (for all $\nu$), Bond et al. (1991) obtained:
\begin{equation}
\nu f(\nu )=\left( \frac{\nu }{2\pi }\right) ^{\frac{1}{2}}\exp (-\frac{\nu}{2})
\label{eq:bond}
\end{equation}   
In the case of a nonspherical collapse, the shape of the barrier is no longer a constant and moreover 
it depends on mass (Del Popolo \& Gambera 1998; SMT). As shown by ST, for a given barrier shape, $B(S)$, where $S\equiv S_{\ast }\left( \frac{\sigma }{\sigma _{\ast }}\right)^{2}=\frac{S_{\ast }}{\nu}$ and $\sigma _{\ast }=\sqrt{S_{\ast }}=\delta _{co}$, the first crossing distribution is well approximated by:
\begin{equation}
f(S)dS=|T(S)|\exp (-\frac{B(S)^{2}}{2S})\frac{dS/S}{\sqrt{2\pi S}}
\label{eq:distrib}
\end{equation}   
where $T(S)$ is the sum of the first few terms in the Taylor expansion of $B(S)$:
\begin{equation}
T(S)=\sum_{n=0}^{5}\frac{(-S)^{n}}{n!}\frac{\partial ^{n}B(S)}{\partial S^{n}}
\label{eq:expans}
\end{equation}   
The previous Eq. ~(\ref{eq:distrib}),(\ref{eq:expans}), reduce to Eq. ~(\ref{eq:bond}) for constant 
barriers. In the case of the ellipsoidal barrier shape given in ST:
\begin{equation}
B(\sigma ^{2},z)=\sqrt{a}\delta _{c}(z)\left[ 1+\frac{\beta }{\left( a\nu \right) ^{\alpha }}\right] 
\end{equation}
where $a=0.707$, $\delta_{\rm c}(z)=1.686 (1+z)$, $\beta \simeq 0.485$ and $\alpha \simeq 0.615$,   
Eqs. ~(\ref{eq:distrib}),(\ref{eq:expans}), give, after truncating the expansion at $n=5$ (see ST):
\begin{equation}
f(\nu )d\nu=A \left( 1+\frac{0.094}{\left( a\nu \right) ^{0.6}}\right) \sqrt{\frac{a\nu }{2\pi }}\exp{\{-a\nu \left[ 1+\frac{0.5}{\left( a\nu \right) ^{0.6}}\right] ^{2}/2\}}
\label{eq:sstt}
\end{equation}
with $A \simeq 1$,
which is in good agreement with the fit of the simulated first crossing distribution (ST):
\begin{equation}
f(\nu )d\nu =A_1\left( 1+\frac{1}{\left( a\nu \right) ^{p}}\right) \sqrt{\frac{a\nu }{2\pi }}\exp (-a\nu /2)
\label{eq:ssttt}
\end{equation}
where $p=0.3$, $A_1=0.3222$ and $a=0.707$. 
\footnote{
Note, that Eq. \ref{eq:ssttt} gives a better fit to Eq. \ref{eq:sstt} if $A \simeq 0.3$ and $a \simeq 0.79$. Viceversa a smaller value of $a$ ($a \simeq 0.63$) and $A=1.08$ in Eq. \ref{eq:sstt} gives a better fit to Eq. \ref{eq:ssttt} (with $A_1=0.3222$ and $a=0.707$), which was the one ST used to compare model and data.}


In the case of  
the ``conditional" mass function, ST showed that an approximation 
can be obtained making 
the replacements $B \rightarrow B(s)- B(S)$ and $S \rightarrow s-S$ in Eq. ~(\ref{eq:distrib}), (\ref{eq:expans}), that means:
\begin{equation}
f(s|S)ds=|T(s|S)|\exp (-\frac{\left[ B(s)-B(S)\right] ^{2}}{2(s-S)})\frac{ds/(s-S)}{\sqrt{2\pi (s-S)}}
\label{eq:distribc}
\end{equation}
\begin{equation}
T(s|S)=\sum_{n=0}^{5}\frac{(S-s)^{n}}{n!}\frac{\partial ^{n}\left[ B(s)-B(S)\right] }{\partial s^{n}}
\label{eq:expansc}
\end{equation}
where as previously reported:
%
\begin{equation}
N(m,\delta _{1}|M,\delta _{o})dm\equiv \frac{M}{m}f(m,\delta _{1}|M,\delta _{o})dm
\end{equation}
where $f(m|M)dm=f(s|S)ds$, $\delta_1=\delta(z_1)$ and $\delta_0=\delta(z_0)$ .

Thus, given Eqs. ~(\ref{eq:distrib})-(\ref{eq:expans}), (\ref{eq:distribc})-(\ref{eq:expansc}), it is possible to obtain 
both the ``unconditional" and ``conditional" mass function, if the barrier shape and the power spectrum are given. 
In the following, I'll use the barrier obtained in Del Popolo \& Gambera (1998) to get the mass functions, which shall 
be compared with those obtained by SMT, ST and with numerical simulations, for several cosmologies. Since the way the barrier is obtained is described in previous papers (see Del Popolo \& Gambera 1998, 1999, 2000) the reader is referred to those papers for details.  
Assuming that the barrier is proportional to the threshold for the collapse, 
similarly to ST, the barrier can be expressed in the form:
\begin{equation}
B(M)=\delta _{\rm c}=\delta _{\rm co}\left[ 1+
\int_{r_{\rm i}}^{r_{\rm ta}}  \frac{r_{\rm ta} L^2 \cdot {\rm d}r}{G M^3 r^3}%
\right] \simeq \delta _{\rm co} \left[ 
1+\frac{\beta_1}{\nu^{\alpha_1}}
\right]
\label{eq:ma7} 
\end{equation}
where $\delta _{\rm co}=1.68$ is the critical threshold for a spherical model,
$r_{\rm i}$ is the initial radius, $r_{\rm ta}$ is the turn-around radius, 
$L$ the angular momentum, $\alpha_1=0.585$ and $\beta_1=0.46$. 
The angular momentum appearing in Eq. ~(\ref{eq:ma7}) is the total angular momentum acquired by the proto-structure during evolution. In order to calculate $L$, I'll use the same model
as described in Del Popolo \& Gambera (1998, 1999) (more hints on
the model and some of the model limits can be found in
Del Popolo, Ercan \& Gambera (2001)).
The CDM spectrum used in this paper is that of Bardeen et al. (1986)(equation~(G3)), with transfer function:
\begin{equation}
T(k) = \frac{[\ln \left( 1+2.34 q\right)]}{2.34 q}
\cdot [1+3.89q+
(16.1 q)^2+(5.46 q)^3+(6.71)^4]^{-1/4}
%
%
\label{eq:ma5}
\end{equation}
where 
$q=\frac{k\theta^{1/2}}{\Omega_{\rm X} h^2 {\rm Mpc^{-1}}}$.
Here $\theta=\rho_{\rm er}/(1.68 \rho_{\rm \gamma})$
represents the ratio of the energy density in relativistic particles to
that in photons ($\theta=1$ corresponds to photons and three flavors of
relativistic neutrinos).
The power spectrum was normalized to reproduce the observed abundance of rich 
cluster of galaxies (e.g., Bahcal \& Fan 1998).
The barrier given in Eq. ~(\ref{eq:ma7}), differently from that of the spherical collapse is mass dependent. 
A direct comparison of the threshold given in Eq. ~(\ref{eq:ma7}) and that given in SMT (Eq. 4) is  
shown in Fig. 1. The dashed line
represents $\delta_{\rm c}(\nu)$ obtained with the present model, while the
solid line that of SMT. Both models show that the threshold
for collapse decreases with mass, or similarly it increases with $\sigma$ since this quantity is a decreasing function of mass.

In other words, this means that, in order to form structure, more massive peaks must
cross a lower threshold, $\delta_c(\nu)$, with respect to under-dense ones.
At the same time, since the
probability to find high peaks is larger in more dense regions, 
this means that, statistically, in order to form structure, 
peaks in more dense
regions may have a lower value of the threshold, $\delta_c(\nu)$, with respect
to those of under-dense regions.
This is due to
the fact that less massive objects are more influenced by external tides, and
consequently they must be more overdense to collapse by a given time.
In fact, the angular momentum acquired by a shell centred on a peak
in the CDM density distribution is anti-correlated with density: high-density
peaks acquire less angular momentum than low-density peaks
(Hoffman 1986; Ryden 1988).
A larger amount of angular momentum acquired by low-density peaks
(with respect to the high-density ones)
implies that these peaks can more easily resist gravitational collapse and consequently 
it is more difficult for them to form structure.
%
%
Therefore, on small scales, where the shear is statistically greater,
structures need, on average, a higher  density contrast to collapse.
%
%

%
%

Putting Eq. (\ref{eq:ma7}) into Eqs. ~(\ref{eq:distrib})-(\ref{eq:expans}) and truncating the expansion at $n=5$, 
I get the ``unconditional" mass function which can be approximated by:
\begin{equation}
f(\nu )d\nu \simeq 1.21\left( 1+\frac{0.06}{\left( a\nu \right) ^{0.585}}\right) \sqrt{\frac{a\nu }{2\pi }}\exp{\{-a\nu \left[ 1+\frac{0.57}{\left( a\nu \right) ^{0.585}}\right] ^{2}/2\}}
\end{equation}
where $a=0.707$.

In a similar way it is possible to obtain the ``conditional" mass function (putting this time Eq. ~\ref{eq:ma7} 
into Eqs. ~(\ref{eq:distribc})-(\ref{eq:expansc})). 
%
%

\section{The temperature function}

The haloes mass functions of clusters of galaxies 
contains information on the structure formation history of the universe: they are a primary input 
for modeling galaxy formation. The mass function is also a critical ingredient in putting strong constraints on 
cosmological parameters (principally $\Omega_0$ and $\Lambda$). 
Observationally the local mass function has been derived from
measuring masses of individual clusters from galaxy velocity dispersions or 
other optical properties by Bahcall and Cen (1993), Biviano et al. (1993), and
Girardi et al. (1998). However, the estimated virial masses for individual clusters
depend rather strongly on model assumptions.
As argued by Evrard et al. (1997) on the basis of hydrodynamical
N-body simulations, cluster masses may be presently more accurately 
determined from a temperature measurement 
and a mass-temperature relation determined from detailed observations
or numerical modeling.
Thus alternatively, as a well-defined observational quantity,
the X-ray temperature function (XTF) has been measured,
which can be converted to the MF by means of the mass-temperature relation.
Recent observational improvement on the XTF was made 
by Markevitch (1998), Henry (2000),
Blanchard et al. (2000), and Pierpaoli et al. (2001),
using more accurate temperature-measurement results
for each cluster with {\it ASCA} data (Tanaka et al. 1994).

The evolution of the temperature distribution is very sensitive to the adopted model of structure 
formation, and its evolution at moderate redshift is considered a crucial test for cosmological models (Kitayama \& Suto 1996).
The cluster temperature function is defined as:
\begin{equation}
N(T,z)=N(M,z) \frac{d M}{dT}
\end{equation}
While the mass function, $N(M,z)$, gives the mass and distribution of a population of 
evolving clusters, the Jacobian $\frac{d M}{dT}$ describes the physical properties of the single cluster.

Comparison of the predictions of the PS theory with the SCDM and OCDM cosmologies, performed by 
Tozzi \& Governato (1998) and Governato et al. (1999), have shown discrepancies between PS predictions and 
N-body simulations, increasing with increasing $z$.  

In the following, I'll use the mass function modified as 
described in the previous section
and an 
improved form of the M-T relation in order to calculate the mass function.

As previously reported, in order to estimate the multiplicity function of real systems  one needs 
to know the temperature-mass (M-T) relation in
order to transform the mass distribution into the temperature distribution.

Theoretical uncertainty arises in this transformation because the exact
relation between the mass
appearing in the PS expression and the temperature of the intra-cluster
gas is unknown.
Under the standard assumption of the 
Intra-Cluster (IC) gas in hydrostatic equilibrium with the potential well 
of a spherically symmetric, virialized cluster, the IC gas temperature-mass
relation is easily obtained by applying the virial
theorem
and for a flat matter-dominated Universe
it is given by (Evrard 1990, Evrard et al. 1996, Evrard 1997):
\begin{equation}
T=(6.4h^{2/3}keV)\left( \frac M{10^{15}M_{\odot }}\right) ^{2/3}(1+z) 
\label{eq:ma2}
\end{equation}
The assumptions of perfect hydrostatic equilibrium and virialization
are in reality not completely satisfied in the case of clusters. Clusters profile may
depart from isothermality, with slight temperature gradients
throughout the cluster (Komatsu \& Seljak 2001). The X-ray weighted temperature can be slightly
different from the mean mass weighted virial temperature. 
%
%
A noteworthy drawback of previous analyses
has been stressed by Voit \& Donahue (1998) (hereafter V98) 
and Voit (2000) (hereafter V2000). Using the merging-halo formalism of Lacey \& Cole (1993), 
which accounts for the fact that massive clusters accrete matter quasi-continuously, they showed that 
the M-T relation evolves, with time, more modestly than what expected in previous models predicting $T \propto (1+z)$, 
and this evolution is even more modest in open universes. 
Moreover, recent studies have shown that the self-similarity in the M-T relation seems to break at 
some keV (Nevalainen, Markevitch \& Forman (hereafter NMF); Xu, Jin \& Wu 2001). By means of ASCA data, 
using a small sample of 9 clusters (6 at 4 keV and 3 at $\sim 1$ keV), NMF has shown that $M_{\rm tot} \propto T_{\rm X}^{1.79 \pm 0.14}$ for the whole sample, and 
$M_{\rm tot} \propto T_{\rm X}^{3/2}$ excluding the low-temperature clusters. 
Xu, Jin \& Wu (2001) has found $M_{\rm tot} \propto T_{\rm X}^{1.60 \pm 0.04}$ (using the $\beta$ model), and $M_{\rm tot} \propto T_{\rm X}^{1.81 \pm 0.14}$ by means of the Navarro, Frenk \& White (1995) profile.  
Finoguenov, Reiprich \& B\"oeringer (2001), have investigated the T-M relation in the low-mass end finding that $M \propto T^{\sim 2}$, and $M \propto T^{\sim 3/2}$ at the high mass end. This behavior has been attributed to the effect of the formation redshift (Finoguenov, Reiprich \& B\"oeringer 2001) (but see Mathiesen 2001 for a different 
point of view), or to cooling processes (Muanwong et al. 2001) and heating (Bialek, Evrard \& Mohr 2000). 
Afshordi \& Cen (2001) (hereafter AC) have shown that non-sphericity introduces an asymmetric, mass dependent, 
scatter for the M-T relation altering its slope at the low mass end ($T \sim 3$ keV).\\
Clearly this has effects on the final shape of the temperature function. 
In the following, I'll use a modified version of the M-T relation obtained 
improving V98, V2000, to take account of tidal interaction between clusters. This M-T relation is given by:
\begin{equation}
kT \simeq 8 keV \left(\frac{M^{\frac 23}}{10^{15}h^{-1} M_{\odot}}\right)
\frac{
\left[
\frac 1m+\left( \frac{t_\Omega }t\right) ^{\frac 23}
+\frac{K(m,x)}{M^{8/3}}
\right]
}
{
\left[
\frac 1m+\left( \frac{t_\Omega }{t_{0}}\right) ^{\frac 23}
 +\frac{K_0(m,x)}{M_0^{8/3}}
\right]
}
\label{eq:kTT1}
\end{equation}
(see Appendix for a derivation), 
where $t_\Omega =\frac{\pi \Omega _0}{H_o\left( 1-\Omega _0-\Omega _\Lambda \right) ^{\frac 32}}$, 
$m=5/(n+3)$ (being $n$ the spectral index), and:
\begin{eqnarray}
K(m,x)&=&F x \left (m-1\right ) {\it LerchPhi}
(x,1,3m/5+1)-
\nonumber \\
& &
F \left (m-1\right ){\it LerchPhi}(x,1,3m/5)
\end{eqnarray}
where $F$ is defined in the Appendix and the 
{\it LerchPhi} function is defined as follows: 
\begin{equation} 
LerchPhi(z,a,v)=\sum_{n=0}^{\infty} \frac{z^n}{(v+n)^a}
\end{equation}
and where $K_0(m,x)$ indicates that $K(m,x)$ must be calculated assuming $t=t_0$.
\begin{figure}
\label{Fig. 1} 
\resizebox{8cm}{!}{\includegraphics{mc393.f1}}
\resizebox{8cm}{!}{\includegraphics{mc393.f1}}
\parbox[b]{9cm}{\caption{The threshold $\delta_{\rm c}$  as a function of the mass M,
through the peak height $\nu$,
taking account of non-radial motions with the model of this paper
(dashed line) is compared with the result of Sheth, Mo \& Tormen (2001) (solid line), obtained
using an ellipsoidal collapse model.}}
\hspace{0.2cm}
\parbox[b]{9cm}{\caption{The simulations refer to a 
$\Lambda$CDM model ($\Omega_0=0.3$, $\Omega_{\Lambda}=0.7$, $h=0.7$. 
The solid line represents the Sheth, Mo \& Tormen (2001) result while the 
dashed line that of this paper.
Filled triangles, open triangles, open squares, filled hexagons and open hexagons 
show results for $z=0$, $z=0.5$, $z=1$, $z=2$ and $z=4$ in the GIF simulations.}}
\resizebox{8cm}{!}{\includegraphics{mc393.f3}}
\resizebox{8cm}{!}{\includegraphics{mc393.f4}}
\parbox[b]{9cm}{\caption{The unconditional mass function. 
The simulations refer to a SCDM model ($\Omega_0=1$, $\Omega_{\Lambda}=0$, $h=0.5$). 
Data are taken from Tozzi \& Governato (1998): open squares refers to $z=0$; filled squares to $z=0.5$ and filled triangles to $z=1$. The solid line represents the PS prediction for $z=0$, the long dashed that for $z=0.5$ and the short dashed that for $z=1$.}}
\hspace{0.2cm}
\parbox[b]{9cm}{\caption{The unconditional mass function. Same as Fig. 3 but now the lines represents 
the prediction obtained using Eq. (\ref{eq:nmm}).
}}
\end{figure}

Eq. (\ref{eq:kTT1}) accounts for the fact that massive clusters accrete matter quasi-continuously, and takes account of tidal interaction between clusters.
The obtained M-T relation is no longer self-similar, a break in the low mass
end ($T \sim 3-4 {\rm keV}$) of the M-T relation is present.
The behavior of the M-T relation is as usual, $M \propto T^{3/2}$,
at the high mass end, and $M \propto T^{\gamma}$, with a value of
$\gamma>3/2$ in dependence of the chosen cosmology.
Larger values of $\gamma$ are related to open cosmologies, while
$\Lambda$CDM cosmologies give results of the slope intermediate
between the flat case and the open case. 

With the previous prescriptions, I am going to calculate the cumulative temperature 
function, $N(>kT)$, and to compare it to simulations. 

\section{Results}

In this section, I compare the ``unconditional"
and ``conditional" mass
functions with measurements in numerical simulations. The
unconditional mass functions in Fig. 2 are taken from SMT, whereas the
measurements in Figs. 3 and 4 are from Tozzi \&
Governato (1998). The ``conditional" mass functions in Figs. 5 and 6 are
reproduced from ST. The ST and SMT measurements were made in
simulations which are a subset of those which were kindly made
available to the public by the Virgo collaboration.

In Fig. 2, I plot the ``unconditional" mass function, obtained as described in the previous section, for
a $\Lambda$CDM model as a function of the scaled variable $\nu$ 
\footnote{Even if not plotted, the ``unconditional" mass function was calculted for SCDM and OCDM models. The results in these cases is very similar to the $\Lambda$CDM model plotted and the discussion relative to 
this case is valid for the other two}. The solid line in the figure represents the ``unconditional" mass function obtained by SMT, the dashed line the same quantity obtained from the barrier given in Eq. (\ref{eq:ma7}). 
The PS prediction is not plotted since from one hand we are more interested in comparison with the Sheth \& Tormen (1999), SMT model, and on the other hand the comparison with the PS formula was performed in Sheth \& Tormen (1999) who showed that the PS prediction are not in agreement with simulation at both high and low $\nu$.

\begin{figure}
\resizebox{6.4cm}{!}{\includegraphics{mc393.f5}}
\hspace{2cm} 
\resizebox{6.4cm}{!}{\includegraphics{mc393.f6}}
\caption{The conditional mass function for a SCDM model ($\Omega_0=1$, $\Omega_{\Lambda}=0$, $h=0.5$) at a redshift $z=0.5$ (left panel). The solid line represents the ST prediction while the dashed line that of the present paper. 
The symbols show the result of numerical simulations (see ST) for parent haloes with mass in the $16 \leq M/M_{\ast} \leq 32$ range (filled hexagons). The right panel is the same as the left one but for $z=4$.
}
\resizebox{6.4cm}{!}{\includegraphics{mc393.f7}}
\hspace{2cm}
\resizebox{6.4cm}{!}{\includegraphics{mc393.f8}}
\caption{The conditional mass function for a $\Lambda$CDM model ($\Omega_0=0.3$, $\Omega_{\Lambda}=0.7$, $h=0.7$) at a redshift $z=0.5$ (left panel). The solid line represents the ST prediction while the dashed line that of the present paper. The symbols show the result of rescaling the conditional mass functions in the simulations for parent haloes with mass in the range 1-2 $M_{\ast}$ (filled hexagons), and 8-32 $M_{\ast}$ (open triangles) at $z=0$. 
The right panel is the same as the left one but for $z=4$.
}
\resizebox{6.cm}{!}{\includegraphics{mc393.f9}}
\resizebox{6.cm}{!}{\includegraphics{mc393.f10}}
\resizebox{6.cm}{!}{\includegraphics{mc393.f11}}
\caption{The cumulative temperature function and the accretion law. 
The simulations refer to a SCDM model ($\Omega_0=1$, $\Omega_{\Lambda}=0$, $h=0.5$, $z=0$) (left panel). 
The solid line represents the model of this paper for the temperature function, while the 
open squares Tozzi \& Governato (1998) simulations. The dotted lines the PS prediction.
The central panel is the same as the left one but for $z=0.4$. The right panel represents the growth curve
for halos of $5 \times 10^{14} h^{-1} M_{\odot}$. The solid line represents Wechsler et al. (2002) prediction (Eq. \ref{eq:wech}), while the dotted line V98 and V2000 prediction (used in this paper).  
}
\end{figure}

Filled triangles, open triangles, open squares, filled hexagons and open hexagons show respectively results for $z=0$, $z=0.5$, $z=1$, $z=2$ and $z=4$ in the so called GIF simulations, a joint effort of astrophysicists from Germany \& Israel, (see Kauffmann et al. 1999; SMT). 

The simulations refer to 
a flat model with non-zero cosmological constant, $\Lambda$CDM model ($\Omega_0=0.3$,$\Omega_{\Lambda}=0.7$, $h=0.7$). Fig. 2 show that 
the Sheth \& Tormen (1999) model and that of this paper gives a good description of the mass function at all output times, suggesting, as remarked by Sheth \& Tormen (1999) and ST, that the dynamics of the collapse is sensitive to $\nu$, and not to the mass scale.
So, in the excursion set model the ``unconditional" mass function, when expressed as a function of $\nu$, is an universal function of $\nu$, independent of redshift, cosmology or initial power spectrum. 

A comparison between the result of Sheth \& Tormen (1999) and that of this paper shows that the model presented here predicts a slightly larger value of the mass function at $ 0.1\leq \nu \leq 4$, 
and even if not plotted in Fig. 2 \footnote{The situation is similar to that of Fig. 10. In Fig. 2, I plotted only values of $\nu \ge 0.1$ to compare it directly  to Fig. 2 of Sheth \& Tormen (1999)}, a smaller value for small $\nu$ ($\nu<0.1$) (at $\nu=0.01$ there are differences of $\simeq 20 \%$). At large $\nu$, the mass function of this paper is smaller than that of Sheth \& Tormen (1999).  
The good agreement between Sheth \& Tormen (1999) model and that of the present paper is due to the similitude of the 
barriers of the two papers. In both two, the barrier increases with $S$ differently from other models (see Monaco 1997a, b). 
It is interesting to note that the increasing of the barrier with $S$ has
several important consequences and these models have a richer structure
than the constant barrier model. In the case of non-spherical collapse
with  
increasing 
barrier, a small fraction of 
the mass in the universe remains unbound, while for the spherical dynamics, at the given time, all the mass is bound 
up in collapsed objects. Moreover, incorporating the non-spherical collapse 
with increasing 
barrier in the excursion set approach results in a model in which
fragmentation and mergers may occur (ST). If the barrier decreases with
$S$ (Monaco 1997 a,b), this implies that all walks are guaranteed
to cross it and so there is no fragmentation associated with this
barrier shape.

As expressed in Eq. (\ref{eq:universal}), the comoving number density of dark matter haloes of mass $m$ in the interval $dM$ and at redshift $z$ is connected to $\nu f(\nu)$. In the case of a constant barrier, it can be explicitly expressed as (PS):
\begin{equation}
n(m,z)=\sqrt{\frac{2}{\pi }}\frac{\rho }{m^{2}}\frac{\delta _{c}}{\sigma D(z)}|\frac{d\ln \sigma }{d\ln m}|\exp (-\frac{\delta _{c}^{2}}{\sigma ^{2}D(z)})
\end{equation}
where $D(z)$ is the linear growth factor normalized to unity at $z=0$ (Peebles 1993), and as previously reported $\delta _{c}$ is the linearly evolved density contrast of fluctuations that are virializing at $z=0$ ($\delta _{c}$=1.686 for $\Omega_0=1$ and 1.65 for the case $\Omega_0=0.3$). 
In the case of the barrier of Eq. ~(\ref{eq:ma7}), we have:
\begin{equation}
n(m,z)\simeq 1.21 \frac{\overline{\rho}}{m^{2}}\frac{d\log (\nu )}{d\log m} 
\left( 1+\frac{0.06}{\left( a\nu \right) ^{0.585}}\right) \sqrt{\frac{a\nu }{2\pi }}\exp{\{-a\nu \left[ 1+\frac{0.57}{\left( a\nu \right) ^{0.585}}\right] ^{2}/2\}}
\label{eq:nmm}
\end{equation}

In Fig. 3, I compare the mass function for a SCDM model obtained by means of numerical simulations (Tozzi \& Governato 1998) at different redshift (open squares, $z=0$; filled squares, $z=0.5$; filled triangles, $z=1$), with the prediction of the PS formula for the same redshifts (solid line, $z=0$; long dashed line, $z=0.5$; short dashed line, $z=1$).  The plot shows a discrepancy between the numerical simulations and the PS predictions. In particular PS 
underestimates the number density of 
high mass clusters ($M>>M_{\ast}$),  discrepancy that grows at higher redshift. For example, at $z=1$ the number density is underestimated by a factor $\geq 3$ for $M>3 \times 10^{14} M_{\odot}$. The excess of massive clusters at higher redshifts suggests that the cluster mass function for the SCDM model evolves more slowly than the PS mass function. The deficit of low mass haloes ($M<<M_{\ast}$) documented in numerical works, e.g. Carlberg \& Couchman (1998), has been associated with merger events not accounted for by the PS formalism. 
As previously reported, incorporating the non-spherical collapse with increasing 
barrier in the excursion set approach results in a model in which fragmentation and mergers may occur, differently from the PS model.
The PS mass function depends sensitively on $\delta_{\rm c}$ and only changing its value and the $z$ dependence it is possible to fit the numerical simulations with the PS formula (Governato et al. 1999).

Fig. 4 gives the same information of Fig. 3, but now the lines represent the predictions of Eq. ~(\ref{eq:nmm}). The plot shows a good agreement with the numerical simulations. So the new mass function is in much better agreement with simulations in agreement with what seen in Fig. 2. 

As previously described, an approximation to the conditional mass function can be obtained from Eq. (\ref{eq:distribc}), (\ref{eq:expansc}). As stressed by ST, the analytical approximation should be more accurate in the case of high-redshift progenitors of massive parents, and less accurate when the parents are not very massive.
Fig. 5 plots the ``conditional" mass function for a SCDM model ($\Omega_0=1$, $\Omega_{\Lambda}=0$, $h=0.5$) at a redshift $z=0.5$ (left panel) and $z=4$ (right panel). The solid line represents the Sheth \& Tormen prediction while the dashed line that of the present paper. The symbols show parent haloes with mass in the $16 \leq M/M_{\ast} \leq 32$ range (filled hexagons) at $z=0.5$. 
Fig. 5, shows that the excursion set predictions with the non-spherical collapse are in good agreement with simulations, while the spherical collapse disagrees with simulations both at small and high values of $m$ and for different redshifts. A comparison between ST and that of the present paper shows a slight difference between the conditional mass functions, difference which is larger at small $m$. 

The difference between the mass functions reflects the difference, shown in Fig. 1, between the barrier obtained by SMT and that obtained in Del Popolo \& Gambera (1998) and used in this paper (see the following of this section for more insight on this point).

The previous arguments have shown that the excursion set approach with non-spherical collapse gives a good description of the conditional and unconditional mass function. In particular the unconditional mass function is an universal function of $\nu$. In the case of the conditional mass function the situation is different, as shown in Figs. 6.
In Fig. 6, I plot the rescaled ``conditional" mass function for a $\Lambda$CDM model ($\Omega_0=0.3$, $\Omega_{\Lambda}=0.7$, $h=0.7$) at a redshift $z=0.5$ (left panel) and $z=4$ (right panel). The solid line represents the Sheth \& Tormen prediction while the dashed line that of the present paper. The symbols show the result of rescaling the conditional mass functions in the simulations for parent haloes with mass in the range 1-2 $M_{\ast}$ (filled hexagons), and 8-32 $M_{\ast}$ (open triangles) at $z=0$. 
As stressed by ST, the symbols in Fig. 6 do not overlap exactly, at fixed $z$,  
the conditional mass functions for different parent haloes do not rescale exactly, and moreover the mass functions at different output times do not rescale either (the band traced out by symbols at $z=0.5$ is different from that traced out at $z=0.5$). In other words, the conditional mass function is not a universal function of $\nu$. A comparison between the simulations (symbols) and the theoretical predictions, (curves), shows that they can provide a good fit only at high redshift (Fig. 6 (right panel)). The situation, even if not represented, is similar for the SCDM model. 
The reasons of the discrepancy should be the following:
one possibility is the neglect of correlations between scales (Peacock \& Heavens 1990; Bond et al. 1991; ST), the second is that the parametrization of the collapse described in SMT is too simple and consequently the collapse threshold and the barrier is also too simple. 
In the model of this paper the parametrization of the collapse is slightly different from that of SMT. In this paper, the spherical collapse is modified to take account of the tidal interaction between protoclusters (see Del Popolo \& Gambera 1998 and the appendix of this paper).  
With the parametrization of this paper the conditional mass function at
small $\nu$ is lower than SMT (at $\nu=0.01$, the difference is
$\simeq 20 \%$) and the curve gives a better fit of the conditional
mass function. At large $\nu$ there is very little difference between
the two predictions.
So the plots show that changing the barrier shape influences the conditional mass function especially
at small lookback times. However, if the parametrization of the collapse
may have a certain role in final shape of the conditional mass function,
the role of correlations between scales is surely important especially
at small lookback times (ST). In fact the excursion set approach describes
in a reasonably good way the clustering at high redshift, but less accurately at small redshifts. The reason is due to the fact that at large lookback times the largest part of subclumps constitutes a small fraction of the mass of the parent halo. Neglect of correlations between scales introduces a small error since the smoothing scale associated with subclumps is sufficiently different from that of the parent. At smaller lookback times the situation changes, since subclumps and parent are not so well separated and neglecting the correlations introduces a larger error with respect to the previous case (ST).
Before going on, it is important to stress that the conditional mass functions in 
the simulations are not yet sufficiently well determined to make strong statements about whether 
the simulations prefer the SMT or the barrier predictions of the present paper. Higher resolution 
simulations are required to see whether one or the other is more 
accurate.  
Finally, I have studied the cumulative temperature function. The function I obtained and which I am going to describe differs from predictions based upon the PS formula for two reasons: a) the cluster density is obtained through the excursion set approach but for a non-spherical collapse and a non-constant barrier; b) as described in Sect. (3) the M-T relation is different from the usual self-similar relation $T \propto M^{2/3}$.
Fig. 7 represents the cumulative temperature function  for a SCDM model ($\Omega_0=1$, $\Omega_{\Lambda}=0$, $h=0.5$, $z=0$ (left panel) and $z=0.4$ (central panel)), and the accretion law for halos of $5 \times 10^{14} h^{-1} M_{\odot}$ (right panel). 
The solid line represents the model of this paper for the temperature function, while the 
open squares Tozzi \& Governato (1998) simulations. The dotted lines the PS prediction with $\delta_{\rm c}$ rescaled as done in Tozzi \& Governato (1998), to fit the mass function ($\delta_{\rm c}(z)=1.48 \times (1+z)^{-0.06}$). 
From the plots, it is clear that the PS predicts less massive clusters than simulations. The discrepancy increases with increasing $z$ and $T$: at $T \simeq 10 {\rm keV}$ and $z=0$ the discrepancy is less than a factor of 3
while at $T \simeq 10 {\rm keV}$ and $z=0.4$, the discrepancy amounts almost to an order of magnitude. 
This implies that the cumulative temperature function obtained from simulations evolves more slowly than the PS prediction, in the range of $z$ and $T$ studied. The situation becomes worse at larger $z$ and $T$. This means that trying to rule out or accept a cosmological model exclusively on the basis of comparisons between data and the PS formula should be taken with caution.
As Fig. 7 shows, the corrections introduced by using Eq. (\ref{eq:nmm}) for the mass function and Eq. (\ref{eq:kTT1}) for the M-T relation, noteworthy improves the agreement with numerical simulations. The approach proposed is thus much more realible than the PS model. 
The right panel of Fig. 7 compares the accretion law calculated according to V98, V2000 model (used in the calculation of the M-T relation) and that obtained by Wechsler at al. 2002. 
As reported in the Appendix, 
the accretion law used in the calculation of the M-T relation is that obtained by V98 following Lacey \& Cole (1993) prescriptions, namely it is obtained by using the approximation: 
\begin{equation}
\langle \frac{dS}{d\omega }\rangle \approx \frac{S}{\omega}
\label{eq:v98}
\end{equation}
(V98),
where $S\equiv \sigma^2(m)$ and $\omega(t)$ is given in V2000 (see also the Appendix). 
%
%
As described in the Appendix, in the case the fluctuation amplitudes can be described by a power-law, we have that:
\begin{equation}
M \propto \omega^\frac{-3}{n+3}
\end{equation}
Simulations by Wechsler et al. 2002 
show that the  mass growth rate is better described by an exponential form:
\begin{equation}
M(a) \propto exp(-\alpha z), \hspace{0.5cm} a=1/(1+z)
\label{eq:wech}
\end{equation}  
(Wechsler et al. 2002). I compared the two different growth rates in the right panel of Fig. 7, for the case 
of halos of mass $5 \times 10^{14} h^{-1} M_{\odot}$. The solid line plots Eq. (\ref{eq:wech}), while the dashed line represents the prediction of V98, V2000 model. As shown in the plot, Wechsler's prediction is different from V98, V2000 and then from that of this paper. The difference is due to the fact that I used (as in V98 and V2000) the Extended Press-Schechter (EPS) formalism to evaluate the accretion rate,  which can reproduce relative properties of the progenitor halo distribution quite well, but has some difficulties in estimating absolute progenitor masses and the overall conditional mass function (Tormen 1998; Sommerville et al. 2000; Gardner 2001). This is one of the reasons that led many authors to propose alternative expressions for the PS mass function (e.g., SMT, ST). However, note that in the calculation of the cumulative temperature function two factors play a role: the expression for the mass function and the M-T relation. This last introduce a correction of ``second order" in the temperature function when compared with the correction introduced by the new form of mass function used (see also Del Popolo \& Gambera 1999). 

%
%

\section{Conclusions}

In this paper, I calculated the unconditional, conditional mass function and temperature function 
by using the extension of the excursion set model of ST and 
the barrier shape obtained in Del Popolo \& Gambera (1988). I showed that the 
barrier obtained in Del Popolo \& Gambera (1998), which takes account of asphericity and tidal interaction
between proto-haloes, is a better description of the mass functions and temperature function than the spherical collapse and 
is in good agreement with numerical simulations. The results are in good agreement with those obtained 
by ST, only some differences are observed expecially at the low mass end.

The main results of the paper can be summarized as follows: \\
1) the non-constant barrier obtained from the non-spherical collapse in Del Popolo \& Gambera (1998), taking account of the tidal interaction of proto-clusters with neighboring ones, combined with the ST model gives ``unconditional" and ``conditional" mass functions, is in reasonably good agreement with results from numerical cosmological simulations. \\
2) The ``unconditional" and ``conditional" mass functions obtained with the Del Popolo \& Gambera (1998) barrier are slightly different from those obtained by SMT and ST: smaller values at small and large $\nu$, with respect to SMT and ST predictions, and larger values for $0.1 \leq \nu \leq 4$. In the case of the rescaled ``conditional" mass function the discrepancy observed by ST for small lookback time is smaller in the model of this paper at small $\nu$ ($0.01 \leq \nu \leq 1$). The discrepancy with simulations is connected to SMT parametrization of the collapse and not only to the neglect of correlations between scales.\\
3) The mass function in SCDM \footnote{Similar conclusion is valid for the OCDM.} 
is in good agreement with Governato et al. (1999) and Tozzi \& Governato (1998) simulations.\\
4)  The temperature function calculated by means of the mass functions obtained in the present paper together with an improved version of the M-T relation, \footnote{As described in the text, the new M-T relation accounts for the fact that massive clusters accrete matter quasi-continuously, 
and consequently that the M-T relation evolves, with time, more modestly than what expected 
in previous models (top-hat model)} 
and taking account of the tidal interaction with neighboring clusters
%
%
is in good agreement with the simulations of SCDM universes of Tozzi \& Governato (1998) for different redshifts. \\
5) The ST formulae really do work  for different barrier shapes (at least that used in this paper, that introduced in SMT and that of Monaco (1997a,b)).\\
6) The behavior of the ``unconditional" mass function at small masses is similar to that of Sheth \& Tormen (1999), ST, and very different from that proposed by Jenkins et al. (2001) (see ST Fig. 13).

The above considerations show that it is possible to get accurate predictions for a number of statistical quantities associated with the formation and clustering of dark matter haloes by incorporating a non-spherical collapse in the   
excursion set approach. The improvement is probably connected also to the fact that 
incorporating the non-spherical collapse with increasing 
barrier in the excursion set approach results in a model in which fragmentation and mergers may occur, effects important in structure formation.
 


\section{Appendix}

From the simulations and observations, we know that the mass within a specified density contrast is
straightforwardly related to temperature. 
%
%
Two fundamental steps in obtaining the M-T relation are: a) 
to 
approximate cluster formation with the evolution of a spherical 
top-hat perturbation (e.g., Peebles 1993); b) 
to assume that each cluster we see at a given 
redshift $z$ has just reached the moment of virialization, an 
assumption known as the late-formation approximation, then 
$M_{\rm vir} \propto T_{\rm X}^{3/2} \rho_{\rm cr}^{-1/2} 
\Delta_{\rm vir}^{-1/2}$.
Some shortcomings of this approach has been summarized in the previous section (see also Viana \& Liddle 1996; Kitayama \& Suto 1996; Eke et al. 1996). 
The late-formation approximation is a good one for many purposes, but a better one can be obtained in the low-$\Omega$ limit. As can be found in the literature, there are two ways of improving the quoted model. One is to define a formation redshift $z_{\rm f}$ at which a cluster virializes and after the properties of observed clusters at $z$ are obtained by integrating over the appropriate distribution of formation redshifts (Kitayama \& Suto 1996; Viana \& Liddle 1996). The second possibility is that described by V98, V2000. In this approach, the top-hat cluster formation model is substituted by a model of cluster formation from spherically symmetric perturbations with negative radial density gradients. The fact that clusters form gradually, and not instantaneously, is taken into account in the merging-halo formalism of Lacey \& Cole (1993). In hierarchical models for structure formation, the growth of the largest clusters is quasi-continuous since these large objects are so rare that they almost never merge with another cluster of similar size (Lacey \& Cole 1993). So, Lacey \& Cole (1993) approach extends the PS formalism by considering how clusters grow via accretion of smaller virialized objects. 
To start with, I'll write the equation governing the collapse of a density perturbation taking account of 
angular momentum acquisition by proto-structures (see Peebles 1993; Del Popolo \& Gambera 1998, 1999; V2000).
In the model, the radial
acceleration of the particle is: 
\begin{equation}
\frac{dv_r}{dt}=\frac{L^2(r)}{M^2r^3}-g(r)=\frac{L^2(r)}{M^{2} r^3}-\frac{G M }{r^2}  
\label{eq:col}
\end{equation}
Assuming a non-zero cosmological constant Eq. (\ref{eq:col}) 
becomes:
\begin{equation}
\frac{dv_r}{dt}=-\frac{G M }{r^2}+\frac{L^2(r)}{M^{2} r^3}+\frac{\Lambda}{3} r \label{eq:coll}
\end{equation}
(Peebles 1993; Bartlett \& Silk 1993; Lahav 1991; Del Popolo \& Gambera 1998, 1999).
Integrating Eq. (\ref{eq:coll}) we have: 
\begin{equation}
\frac{1}{2}\left( \frac{dr}{dt}\right) ^{2}=\frac{GM}{r}+\int 
\frac{L^{2}}{M^{2}r^{3}}dr+\frac{\Lambda }{6}r^{2}+\epsilon
\label{eq:coll1}
\end{equation}
where the value of the specific binding energy of the shell, $\epsilon$, can be obtained using the condition for turn-around, $\frac{dr}{dt}=0$.

Integrating Eq.(\ref{eq:coll1}), I get:
\begin{equation}
t=\int\frac{dr}{\sqrt{2\left[ \epsilon+\frac{GM}r+\int_{r_{\rm i}}^{r} \frac{L^2}{M^2r^3}dr+\frac{\Lambda}{6}r^2\right] }}
\label{eq:tmppp}
\end{equation}
A particular shell will collapse if: 
\begin{equation}
\epsilon+\frac{GM}{r_0}+\int_{r_{\rm i}}^{r_0} \frac{L^2}{M^2r^3}dr+\frac{\Lambda}{6}r_0^2=0
\end{equation}
(see V2000),
and the shell reaches its maximum radius (turn-around radius, $r_{\rm ta}$) at a time:
\begin{equation}
t_{\rm ta}=\int_{0}^{r_0} \frac{dr}{\sqrt{2\left[ \epsilon+\frac{GM}r+\int_{r_{\rm i}}^{r} \frac{L^2}{M^2r^3}dr+\frac{\Lambda}{6}r^2\right]}}
\label{eq:tmp}
\end{equation}
Eq. (\ref{eq:tmp}) can be written in an equivalent form (see Del Popolo \& Gambera 1998, 1999; Bartlett \& Silk 1993), as:
\begin{eqnarray}
t_{ta}&=&\int_{0}^{r_{ta}}\frac{dr}{\sqrt{2\left[ GM\left( \frac{1}{r}-\frac{1}{r_{ta}}\right) +\int_{r_{ta}}^{r}\frac{L^{2}}{M^{2}r^{3}}dr+\frac{\Lambda}{6}(r^2-r^2_{\rm ta})\right] }}=
\nonumber \\
& &
2H_0^{-1}\Omega_0 ^{-1/2}\xi ^{3/2}\int_{0}^{1}
\frac{y^{2}dy}{\sqrt{1+
\left[ \frac{9\xi ^{9}}{4\pi ^{2}\rho ^{2}r_{ta}^{10}\Omega H^{2}}\int_{1}^{y^{2}r_{ta}}\frac{L^{2}}{y^{5}}dy\right]y^2
+\left( \frac{\Omega _{\Lambda }}{\Omega _{0}}\right) \xi ^{3}y^{6}-\left[ 1+\left( \frac{\Omega _{\Lambda }}{\Omega _{0}}\right) \xi ^{3}\right] y^{2}}}
\label{eq:tmp1}
\end{eqnarray}
%
%
%
where
$\xi$ is a parameter given by $ \xi=r_{ta}/x_{1}$, $r_{ta}$ being
the radius of the turn-around epoch, while $x_{1}$ is defined
by the relation
$M=4 \pi \rho_{\rm b} x^{3}_{1}/3$, 
$\Omega_0=\frac{8 \pi G \rho_{\rm b}}{3 H_0^2}$ and
$\Omega_{\Lambda}=\frac{\Lambda}{3 H_0^2}=1-\Omega_0$, and $\rho_{\rm b}$ the critical density.
The shell collapses at $t_{\rm c}=2 t_{\rm ta}$.
In the case $L=0$, Eq. (\ref{eq:tmp}) can be analytically integrated and inverted to get $\epsilon(t)$:
\begin{equation}
\epsilon(t)=\frac12 \left(\frac{2 \pi G M}{t}\right)^{2/3}=
\frac12 \left(\frac{2 \pi G M}{t_{\Omega}}\right)^{2/3} \left(\frac{t_{\Omega}}{t}\right)^{2/3}=
\frac12 \left(\frac{2 \pi G M}{t_{\Omega}}\right)^{2/3} (x-1)
\label{eq:eps}
\end{equation}
where we have defined $t_\Omega =\frac{\pi \Omega _0}{H_o\left( 1-\Omega _0-\Omega _\Lambda \right) ^{\frac 32}}$ and $x=1+(\frac{t_{\Omega}}{t})^{2/3}$ which is connected to mass by $M=M_{\rm 0} x^{-3 m/5}$, where $m=5/(n+3)$ and $n$ is the usual power-law perturbation index (V2000). 

Integrating with respect to mass $\epsilon(t)$, and diving by mass, one gets, as shown by V2000:
\begin{equation}
\frac{E}{M}=-\frac{\int \epsilon dM}{M}=
\frac{3m}{10(m-1)}\left( \frac{2\pi G}{t_\Omega }\right) ^{\frac 23}M^{\frac 23}\left[ \frac 1m+\left( \frac{t_\Omega }t\right) ^{\frac 23}\right]
\end{equation}
In the case $L \neq 0$, it is numerically possible to integrate and invert Eq. (\ref{eq:tmp}) finding the specific energy,
which we can indicate with $\epsilon_{\rm L}(t)$ and finally:
\begin{equation}
\frac{E}{M}=-\frac35 \frac{m}{m-1}\epsilon_{\rm L}(t) 
\end{equation}
An approximate relation for $\epsilon_{\rm L}(t)$, in an Einstein-de Sitter Universe can be obtained as follows.
It is possible to obtain the turn-around radius, $r_{\rm ta}$ by solving Eq. (\ref{eq:tmp1}) (with $\Lambda=0$) for a given mass and a given epoch of interest. This is related to the binding energy of the shell enclosing the mass $M$ by Eq. (\ref{eq:coll1}) with $\dot r=0$. 
In turn, the binding energy is uniquely given by the linear overdensity $\delta_{\rm i}$ at some arbitrary early time.
We may now use the linear theory to ``propagate" the overdensity to the ``chosen" time to find the linear overdensity at turn-around, $\delta_{\rm c}$.
Using the relation between $v$ and $\delta_{\rm i}$ for the growing mode (Peebles 1980) in Eq. (\ref{eq:coll1}), with $\Lambda=0$, at an early time, it is possible to connect $\epsilon$ and $\delta_c$ (see Bartlett \& Silk 1993)):
\begin{equation}
\epsilon=\frac56 \Omega_0 H_0^2 x_1^2(\delta_{\rm i}/a_{\rm i})=\frac12 \left(
\frac{2 \pi G M}{t}
\right)^{2/3}
\label{eq:epsil}
\end{equation}
Using the formula for $\delta_{\rm c}$ given in Del Popolo \& Gambera (1999, Eq. 14), Del Popolo \& Gambera (2000, Eq. 6), is possible to write:
\begin{equation}
\epsilon_{\rm L}=-\frac12 \left(\frac{2 \pi G M}{t}
\right)^{2/3}\delta_{\rm c}
%
%
=
\frac12 \left(\frac{2 \pi G M}{t}\right)^{2/3}
\left[
1+\frac{r_{\rm ta}}{G M^3} \int_0^{r} \frac{L^2 dr}{r^3}\right]
\label{eq:epsil1}
\end{equation}
which reduces to Eq. (\ref{eq:epsil}) when $L \rightarrow 0$.

Eq. (\ref{eq:epsil1}) can also be written as:
\begin{equation}
\epsilon_{\rm L}=-\frac12 \left(\frac{2 \pi G M}{t_{\Omega}}\right)^{2/3}(x-1)
\left[
1+\frac{2^{7/3} \pi^{2/3} \xi \rho_{\rm b}^{2/3}}{3^{2/3} H^2 \Omega M^{8/3}} \frac{1}{x-1}
\int_0^{r} \frac{L^2 dr}{r^3}\right]
\label{eq:ener1}
\end{equation}
Defining $M=M_{\rm 0} x^{-3 m/5}$ and 
\begin{equation}
F=\frac{2^{7/3} \pi^{2/3} \xi \rho_{\rm b}^{2/3}}{3^{2/3} H^2 \Omega} 
\int_0^r \frac{L^2 dr}{r^3}
\end{equation}
Eq. (\ref{eq:ener1}) becomes:
\begin{equation}
\epsilon_{\rm L}=-\frac12 \left( \frac{2 \pi G}{t_{\Omega}}\right)^{\frac 23}M^{2/3}
\left[ \left( \frac M{M_o}\right) ^{-\frac 5{3m}}-1\right] \left[ 1+\frac F{M^{\frac 83}}\frac 1{\left( \frac M{M_o}\right) ^{-\frac 5{3m}}-1}\right]  
\label{eq:ener2}
\end{equation}
Integrating with respect mass, and dividing again by $M$, I get:
\begin{equation}
\frac{E}{M}=-\frac{\int \epsilon_{\rm L} dM}{M}=
\frac{3m}{10(m-1)}\left( \frac{2\pi G}{t_\Omega }\right) ^{\frac 23}M^{\frac 23}
\left[
\frac 1m+\left( \frac{t_\Omega }t\right) ^{\frac 23}
+\frac{K(m,x)}{M^{8/3}}
\right]
\label{eq:em} 
\end{equation}
where
\begin{eqnarray}
K(m,x)&=&F x \left (m-1\right ){\it LerchPhi
}(x,1,3m/5+1)-
\nonumber \\
& &
\left (m-1\right )F{\it LerchPhi}(x,1,3m/5)
\end{eqnarray}
where the {\it LerchPhi} function is defined as follows: 
\begin{equation} 
LerchPhi(z,a,v)=\sum_{n=0}^{\infty} \frac{z^n}{(v+n)^a}
\end{equation}
%
\footnote{This definition is valid for $abs(z) < 1$. By analytic continuation, it is extended to 
the whole complex z-plane, for each value of a.
If the coefficients of the series representation of a hypergeometric function are rational functions of the summation indices, then the hypergeometric function can be expressed as a linear sum of Lerch Phi functions. 
Reference: A. Erdelyi, 1953, Higher Transcendetal Functions, Volume 1, chapter 1, section 11.} 
%
If $K=0$, Eq. (\ref{eq:em}) reduces to Eq. (10) of V2000. As stressed by V2000, some factors give rise to an higher value of $E/M$ with respect the case of the late-formation value. The $m/(m-1)$ value which accounts for the effect of early infall. The $1/m$ value in the square bracket of Eq. (\ref{eq:em}) which accounts for the cessation of cluster formation when $t>>t_{\rm \Omega}$. Finally in Eq. (\ref{eq:em}) a new term is present, which comes from the tidal interaction.
In order to obtain an expression for the kinetic energy, starting from $E/M$, I use the virial theorem with the surface pressure term correction as in V2000 and 
utilizing the usual relation:
\begin{equation}
\langle K \rangle= \frac{3 \beta M k T}{2 \mu m_{\rm p}}
\label {eq:conn} 
\end{equation}
(AC),
where $k$ is the Boltzmann constant, $\mu=0.59$ is the mean molecular weight, $m_{\rm p}$ the proton mass and $\beta=\frac{\sigma_{\rm v}^2}{kT/\mu m_{\rm p}}$, being $\sigma_{\rm v}$ the mass-weighted mean velocity dispersion of dark matter particles.
In this way, I finally get: 
\begin{equation}
kT=\frac 25a\frac{\mu m_p}{2\beta} \frac m{m-1}\left( \frac{2\pi G}{t_\Omega }\right) ^{\frac 23}M^{\frac 23}
\left[
\frac 1m+\left( \frac{t_\Omega }t\right) ^{\frac 23}
+\frac{K(m,x)}{M_0^{8/3}}
\right]
\label{eq:kT}
\end{equation}
where $a=\frac{\overline{\rho}}{2 \rho(r_{\rm vir})-\overline{\rho}}$ is the ratio between kinetic and total energy (V2000) \footnote{This term comes from the fact I used a modified version of the virial theorem in order to include a
surface pressure term (see also V2000, AC). This correction is due to the fact that
at the virial radius $r_{\rm vir}$ the density is non-zero and this requires
a surface pressure term to be included in the virial theorem
(Carlberg, Yee \& Ellingson 1997) (the existence of this confining pressure
is usually not accounted for in the top-hat collapse model)}.
Using the relation $\Delta_{\rm vir}=\frac{8 \pi^2}{H t^2}$ (see V2000), and in the early-time limit: ($t<<t_{\Omega}$), Eq. (\ref{eq:kT}), reduces to:
\begin{equation}
kT=\frac 25 \frac m{m-1}a\frac{\mu m_p}{2\beta}  G M^{\frac 23}
\left(
\frac{4 \pi}{3} \rho_{\rm b} \Delta_{\rm vir}
\right)^{1/3}
\label{eq:kT1}
\end{equation}
which, in the case $n \sim -2$, $a \sim 2$ is identical 
to the late-formation formula, described in V2000 (see their Eq. (8)).
Normalizing Eq. (\ref{eq:kT}) similarly to V2000, I get:
\begin{equation}
kT \simeq 8 keV \left(\frac{M^{\frac 23}}{10^{15}h^{-1} M_{\odot}}\right)
\frac{
\left[
\frac 1m+\left( \frac{t_\Omega }t\right) ^{\frac 23}
+\frac{K(m,x)}{M^{8/3}}
\right]
}
{
\left[
\frac 1m+\left( \frac{t_\Omega }{t_{0}}\right) ^{\frac 23}
 +\frac{K_0(m,x)}{M_0^{8/3}}
\right]
}
\label{eq:kT1}
\end{equation}
where $K_0(m,x)$ indicates that $K(m,x)$ must be calculated assuming $t=t_0$

\end{document}